\documentclass[aps,pra,twocolumn,superscriptaddress,showpacs,showkeys,amsmath,amssymb]{revtex4}

\usepackage{amsfonts}
\usepackage{amssymb,amsmath}
\usepackage{mathrsfs}
\usepackage{latexsym}
\usepackage{amsmath}
\usepackage[cp1251]{inputenc}
\usepackage{graphicx}
\usepackage{dcolumn}
\usepackage{bm}
\usepackage{color}

\RequirePackage{ifthen}
\RequirePackage[pdfstartview=FitH]{hyperref}
\begin{document}

    \title{Mean-field study of repulsive 2D and 3D Bose polarons}
    \author{O.~Hryhorchak}
    \affiliation{Department for Theoretical Physics, Ivan Franko National University of Lviv, 12 Drahomanov Str., Lviv, Ukraine}
    \author{G.~Panochko}
    \affiliation{Department of Optoelectronics and Information Technologies, Ivan Franko National University of Lviv, 107 Tarnavskyj Str., Lviv, Ukraine}
    \author{V.~Pastukhov\footnote{e-mail: volodyapastukhov@gmail.com}}
    \affiliation{Department for Theoretical Physics, Ivan Franko National University of Lviv, 12 Drahomanov Str., Lviv, Ukraine}

    \date{\today}


    \begin{abstract}
    	The detailed mean-field treatment of the Bose polaron problem in two and three dimensions is presented. Particularly, assuming that impurity is immersed in the dilute Bose gas and interacts with bosons via the hard-sphere two-body potential, we calculate the low-momentum parameters of its spectrum, namely, the binding energy and the effective mass. The limits of applicability of the mean-field approach to a problem of mobile impurity in Bose-Einstein condensates are discussed by comparing our results to the Monte Carlo simulations data.
    \end{abstract}

    \maketitle

\section{Introduction}
\label{sec1}
\setcounter{equation}{0}
It always captures our imagination when simple and physically clear arguments are used to get more insight in a problem. If these arguments are additionally supported by the comparatively simple calculations which correctly explain an experimental situation or reproduce (at least qualitatively) results of essentially exact numerical methods we are twice happily. One of such an effective tool in context of many-body physics is the mean field (MF) approximation, which in its different realizations can describe on qualitative level a variety of the essentially quantum phenomena such as magnetism, superfluidity and superconductivity. The MF approximation in the theory of Bose systems, that is usually associated with Gross \cite{Gross_61} and Pitaevskii \cite{Pitaevskii_61} on their study of the quantized vortices in dilute gases, has found its application, after the realization of the Bose-Einstein condensation of alkali atoms more than two decades ago, in the description \cite{Dalfovo_etal} of ultracold trapped quantum gases.

Besides of explicit accounting for the external trapping potential and description of topologically non-trivial objects in dilute Bose condensates in low dimensions or restricted geometries, the MF was shown \cite{Astrakharchik_04} to be useful in the Bose polaron problem. Being able to describe both the self-localization phenomenon \cite{Cucchietti_06,Kalas_06,Sacha_06,Bruderer_08,Roberts_09,Blinova_13} and translation-invariant \cite{Gross} states of impurity by an appropriate choice of the wave function, the MF approximation was demonstrated \cite{Volosniev_17,Pastukhov_3BIBP,Smith} to be quite accurate analytical tool in the problem of one dimension (1D) Bose polaron. In recent years a single impurity atom immersed in Bose condensates have attracted much attention not least because of the experimental realization of 3D \cite{Jorgensen,Hu} Bose polarons. Equilibrium properties of these system in 3D are explored theoretically at zero \cite{Novikov_09,Novikov_10,Rath_13,Shashi,Li_14,Christensen_15,Grusdt_15,Vlietinck_15,Pena_Ardila_15,Shchadilova,Pena_Ardila_16,GSSD,Pena_Ardila_19} and finite \cite{Levinsen_17,Guenther,Bosepolaron_D,Field} temperatures, and the dynamics is well understood \cite{Volosniev_15,Shchadilova_16,Lampo_18,Lausch,Drescher,Liu,Boyanovsky}. A characteristic feature of the many-body physics in high dimensions is the presence of few-particle effects, fingerprints of which can be detected \cite{Zinner,Levinsen_15,Sun,Yoshida,Naidon} in the Bose polaron behavior. Although the experimental realization of 2D Bose polarons is lacking (in part because of breakdown \cite{Pastukhov_2DBP} of the quasiparticle picture at any finite temperatures), theoretical efforts in this direction have been made \cite{Grusdt_16,Grusdt_Fleischhauer_16,Pena_Ardila_Pohl}. A more deep insight in the ground-state properties of impurity immersed in 2D Bose condensates, however, can be obtained by means of Monte Carlo (MC) simulations, which were recently reported by two groups \cite{Akaturk,Pena_Ardila_Astrakharchik_19}.

It was shown in our recent article \cite{Panochko_19} that application of the MF approximation to the problem of 1D Bose polaron leads to the reasonable low-momentum impurity spectrum. A quite satisfactory coincidence with the results of MC simulations \cite{Parisi_17,Grusdt_17} were found for an impurity immersed both in the system of weakly-interacting bosons and in the Tonks-Girardeau gas. The objective of present article is to extend our previous MF analysis to the Bose polaron problem in higher dimensions. It should be noted that the utilized here extended MF approach is not equivalent to the perturbation theory \cite{Novikov_09,Grusdt_Demler_15,Christensen_15,Panochko_17,Pastukhov_2DBP} even at weak boson-impurity interactions, but it is the simplest analytical method that provides a non-perturbative predictions in the strong-coupling limit. Furthermore, being combined with the local density approximation, the MF gives a qualitative correct insight in a behavior of impurities in the systems for which the consistent microscopic description is lacking.

\section{Model}
\subsection{Basic equations}
The system to be discussed is a single impurity atom immersed in the $D$-dimensional (two- and three-dimensional) Bose environment at absolute zero. It is assumed the thermodynamic limit, i.e., the whole system is loaded in large volume $L^D$ with periodic boundary conditions. Particularly, we are going to explore properties of a semi-phenomenological model with the Hamiltonian
\begin{eqnarray}\label{H}
H=-\frac{\hbar^2}{2m_I}\frac{\partial^2}{\partial {\bf r}_I^2}+H_B+\int d{\bf r}\,\Phi(|{\bf r}_I-{\bf r}|)n({\bf r}),
\end{eqnarray}
where the first term is kinetic energy of impurity with position ${\bf r}_I$ and mass $m_I$. The Hamiltonian of Bose subsystem 
\begin{eqnarray}\label{H_B}
H_B=\int d{\bf r}\left\{-\frac{\hbar^2}{2m}\psi^+({\bf r})\nabla^2\psi({\bf r})+\mathcal{E}[n]\right\},
\end{eqnarray}
goes back to seminal work of Landau \cite{Landau} on his theory of superfluidity of liquid $^{4}$He, describes $N$ bosons of mass $m$ with $\mathcal{E}[n]$ being the normal-ordered energy density of the uniform system at rest. The field operators $\psi^+({\bf r})$, $\psi({\bf r})$ obey standard bosonic commutation relations $[\psi({\bf r}),\psi^+({\bf r}')]=\delta({\bf r}-{\bf r}')$ and $[\psi({\bf r}),\psi({\bf r}')]=0$, while $n({\bf r})=\psi^+({\bf r})\psi({\bf r})$ denotes the local density of bosons. The third term in Hamiltonian (\ref{H}) stands for the Bose-system-impurity interaction with hard-sphere (hard-disk in 2D) two-body potential
\begin{eqnarray}\label{Phi}
\Phi(r)=\left\{\begin{array}{c}
\infty, \ \ r\le a_I\\
0, \ \  r>a_I
\end{array}\right..
\end{eqnarray} 

The non-commutativity of the first and third terms in (\ref{H}) makes the further consideration quite cumbersome, therefore, in order to overcome this difficulty we perform the Lee-Low-Pines \cite{LLP} transformation, $H_U=U^{+}HU$, originally proposed in the polaron problem, where $U=\exp\left\{i{\bf r}_I({\bf p}-{\bf P})/\hbar\right\}$ with operator ${\bf P}=\int d {\bf r}\, \psi^+({\bf r})(-i\hbar\nabla)\psi({\bf r})$ that denotes the momentum carrying by Bose particles. The unitary-transformed Hamiltonian
\begin{eqnarray}\label{H_prime}
H_U=\frac{p^2}{2m_I}+H_B+\int d{\bf r}\,\Phi(r)n({\bf r})+\Delta H_U,
\end{eqnarray}
commutes with $-i\hbar\frac{\partial}{\partial {\bf r}_I}$, which eigenvalue can be chosen arbitrary (zero in our case), but contains additional terms
\begin{eqnarray}\label{Delta_H}  
\Delta H_U=-\frac{{\bf pP}}{m_I} +\frac{{\bf P}^2}{2m_I}.
\end{eqnarray}
It is easy to show that the expectation value of total momentum $-i\hbar\frac{\partial}{\partial {\bf r}_I}+{\bf P}$ of the system, which we associate with impurity motion, is equal to $\bf p$.
In the following we mainly focus on properties of almost motionless impurity, i.e. ${\bf p}\simeq 0$, which means that the average counterflow of bosons $\langle{\bf P}\rangle$ is also almost zero. Note, however, that because of quantum effects $\langle{\bf P}^2\rangle$ does not disappear even when momentum of impurity is exactly zero. 

The time evolution of the field operator can be deduced by using the standard quantum-mechanical prescription
\begin{eqnarray}\label{psi_t}
	i\hbar \frac{\partial}{\partial t}\psi({\bf r},t)=[\psi({\bf r},t), H_{U}]
	=-\frac{\hbar^2\nabla^2}{2m_r}\psi({\bf r},t)\nonumber\\
	+\Phi(r)\psi({\bf r},t)+\mathcal{E}'[n]\psi({\bf r},t)+\frac{i\hbar}{m_I}({\bf p}-{\bf P})\nabla\psi({\bf r},t),
\end{eqnarray}
where $m_r=m_Im/(m_I+m)$ is the reduced mass and $\mathcal{E}'[n]$ represents (the normal-ordered) derivative of the energy density with respect to $n$. It should be noted that till now our consideration is the exact one, and no approximations were made. Indeed, by performing the Lee-Low-Pines transformation we get rid of the explicit dependence on the impurity position ${\bf r}_I$ in Hamiltonian $H_U$ (formally the impurity is now placed at the origin). The price to pay for these simplifications is that the unitary-transformed Hamiltonian necessarily includes extra term $\Delta H_U$, where the second term involves additional kinematic interaction between bosons.

\subsection{MF approximation for motionless impurity}
The mean-field ansatz suggests \cite{Pitaevskii_Stringari} that the field operator $\psi({\bf r},t)$ in Eq.~(\ref{psi_t}) can be replaced by complex function  $e^{-i\mu t/\hbar}\phi({\bf r})$ with $\mu$ being the chemical potential that fixes the average density of the system. In general, the obtained MF equation is very complicated and can be exactly solved only in 1D \cite{Panochko_19}, but for motionless impurity ${\bf p}=0$ the wave function $\phi({\bf r})$ is determined by much more simple equation
\begin{eqnarray}\label{phi}
-\frac{\hbar^2\nabla^2}{2m_r}\phi({\bf r})
+\Phi(r)\phi({\bf r})+\mathcal{E}'[n]\phi({\bf r})=\mu\phi({\bf r}),
\end{eqnarray}
and can be chosen to be real-valued. The latter condition automatically rejects the topologically non-trivial solutions \cite{Braz} of Eq.~(\ref{psi_t}). Formally, $n({\bf r})=|\phi({\bf r})|^2$ represents the density profile of bosons in the external potential of impurity, which in turn, placed at ${\bf r}_I=0$. When $\Phi(r)=0$, the solution of Eq.~(\ref{phi}) is obvious $\phi({\bf r})=\textrm{const}$. Importantly, the quantum-mechanical expectation value of the local density operator $\langle\psi^+({\bf r})\psi({\bf r})\rangle$, where $\langle\ldots\rangle$ necessary contains an averaging both over the bosonic and the impurity states, is always constant and equal to $\bar{n}=\frac{1}{L^D}\int d{\bf r} n({\bf r})$. This is a direct consequence of the continuous translation symmetry of the model. However, keeping in mind Eq.~(\ref{phi}), it is convenient to refer to quantity $n({\bf r})$ as a density profile of bosons.

Having calculated $n({\bf r})$, we are in position to obtain the energy of the system `Bose particles + impurity'
\begin{eqnarray}\label{E}
E=\mu N+\int d{\bf r}\left\{\mathcal{E}[n]-n\mathcal{E}'[n]\right\}.
\end{eqnarray}
Then, subtracting the MF energy of `pure' bosons from $E$ we get the impurity binding energy $\varepsilon_I$.

Because the solution of Eq.~(\ref{phi}) is a non-uniform in the presence of an impurity, it is also instructive to calculate the quasiparticle residue, which is determined by the modulus squared of the wave-function overlap, $Z=\left|\int d{\bf r}_I\langle \Psi_0|\Psi\rangle\right|^2$. Here $|\Psi\rangle=\frac{\left(\tilde{b}^+_0\right)^N}{\sqrt{L^{D}N!}}|\textrm{vac}\rangle$ and
$|\Psi_0\rangle=\frac{\left(b^+_0\right)^N}{\sqrt{L^{D}N!}}|\textrm{vac}\rangle$ are the transformed MF ground-state wave functions of `$N$ bosons + impurity' with zero and non-zero $\Phi(r)$, respectively. Here $|\textrm{vac}\rangle$ is the normalized vacuum state, and bosonic creation operators are related to the field operator in conventional way
\begin{eqnarray}\label{b^+}
b^+_0=\frac{1}{\sqrt{L^{D}}}\int d{\bf r}\psi^+({\bf r}), \  \tilde{b}^+_0=\frac{1}{\sqrt{N}}\int d{\bf r}\phi({\bf r})\psi^+({\bf r}),
\end{eqnarray}
where $\phi({\bf r})$ is the solution of Eq.~(\ref{phi}). Explicit calculations of the overlap then yield
\begin{eqnarray}
\int d{\bf r}_I\langle \Psi_0|\Psi\rangle=\left\{\frac{1}{\sqrt{NL^D}}\int d{\bf r}\phi({\bf r})\right\}^N,
\end{eqnarray}
assuming that $\phi({\bf r})$ is real function we finally obtain the quasiparticle residue in thermodynamic limit $N\gg 1$
\begin{eqnarray}\label{Z}
Z=\exp \left\{-\int d{\bf r}\left(\sqrt{n}-\sqrt{n_{\infty}}\right)^2\right\},
\end{eqnarray}
where $n_{\infty}=n(r\to\infty)$. Equations (\ref{b^+}) readily provide the meaning of function $\phi({\bf r})$. The lowest single-particle energy level of bosons described by Hamiltonian $H_U$ with weak inter-particle interaction and $\Phi(r)=0$ is constant $1/\sqrt{L^D}$, while $\phi({\bf r})$ should be treated as the transformed (unnormalized) wave function of a single boson in the presence of impurity located at the origin. The weak repulsive potential among bosons is very important, because it automatically rejects any collapsed BEC states \cite{Panochko_Pastukhov}, when all non-interacting Bose particles simultaneously form bound states with impurity. It worth noting that the impurity also slightly deforms the excitation spectrum of Bose system, which can be calculated by means of the Bogoliubov-de Gennes formalism \cite{Boudjemaa,Takahashi}.

\subsection{Effects of slow impurity motion}
At small but non-zero ${\bf p}$ the impurity energy increases parabolically
\begin{eqnarray}\label{varepsilon_p}
\varepsilon_I(p)=\varepsilon_I+\frac{p^2}{2m^*_I}+\mathcal{O}(p^4),
\end{eqnarray}
where $m^*_I$ is the effective mass which takes into account the interaction with Bose medium. It is easy to show by using naive speculations that the MF effective mass is always larger than `bare' impurity mass. Indeed, for the repulsive boson-impurity interaction the moving particle has to push apart the surrounding bosons which leads to effective increase of its inertial mass. For the attractive boson-impurity potential, on the other hand, one may think that bosons stick to impurity providing that it gains some additional mass. In both cases, however, the motion of impurity causes the non-zero average bosonic flow $\langle {\bf P} \rangle$, which necessary decreases the energy of the whole system. Actually, the occurrence of the directed motion of Bose particles is responsible for the value of the effective mass
\begin{eqnarray}\label{m_eff}
\frac{m_I}{m^*_I}=1-\frac{1}{D}\left(\frac{\partial}{\partial {\bf p}}\langle {\bf P} \rangle\right)_{{\bf p}=0}.
\end{eqnarray}
So, in order to calculate the average momentum of Bose system we have to solve the Gross-Pitaevskii-like equation (\ref{psi_t}) with non-zero ${\bf p}$. The stationary solution are most simply written in exponential form $e^{-i\mu t/\hbar}|\phi({\bf r})|e^{i\Theta({\bf r})}$, where the phase 
\begin{eqnarray}\label{eq_Theta}
\hbar\nabla\{ n({\bf r})\nabla \Theta({\bf r})\}=\frac{m_r}{m_I}({\bf p}-\langle {\bf P}\rangle)\nabla n({\bf r}),
\end{eqnarray}
satisfies linear equation for any magnitude of ${\bf p}$, while the squared amplitude $n({\bf r})$ of the wave function is always determined by the nonlinear one. For the calculations of the effective mass, however, we only need to know the leading-order small-${\bf p}$ behavior of phase field 
\begin{eqnarray}\label{Theta_simp}
\Theta({\bf r})|_{{\bf p}\to 0}\to\frac{m_r}{m^*_I}\frac{{\bf pr}}{\hbar}\theta(r),
\end{eqnarray}
where ${\bf p}/{m^*_I}=({\bf p}-\langle {\bf P}\rangle_{{\bf p}\to 0})/{m_I}$, and $\theta(r)$ is spherically symmetric function. Substitution of ansatz (\ref{Theta_simp}) in Eq.~(\ref{eq_Theta}), where $n(r)$ is the spherically symmetric solution of Eq.~(\ref{phi}), leads to ordinary second-order differential equation
\begin{eqnarray}\label{eq_theta}
r\frac{d}{dr}\left(n\frac{d\theta}{dr}\right)+Dn\frac{d\theta}{dr}+\frac{d\left(n\theta\right)}{dr}=\frac{dn}{dr},
\end{eqnarray}
The above equation is readily solved analytically in one-dimensional case ($D=1$), while in higher dimensions we can only obtain numerical solutions. But behavior of function $\theta(r)$ at large distances can be evaluated, $\theta(r)\sim A_D/r^D$, because typically local density of bosons $n(r)$ exponentially reaches a constant value at large distances from impurity. In what follows that integral, $\langle {\bf P}\rangle =\hbar \int d{\bf r}n\nabla\Theta$, which determines the average Bose system's momentum  is formally divergent (because the result depends on sequence of integration). Exactly the same situation is realised in classical hydrodynamics \cite{Hydrodynamcs} during the calculations of momentum carrying by the liquid that is disturbed by a rigid body that moves with a constant velocity. Nevertheless this momentum is finite, and in order to calculate it at small $\bf p$ we multiply the both sides of Eq.~(\ref{eq_Theta}) by $-\hbar \Theta({\bf r})/m_r$ and integrate over the volume (area in 2D) of sphere (disk) of large radius $R$
\begin{eqnarray}\label{identity}
\frac{\hbar^2}{m_r}\int_{r\le R} d{\bf r}n(\nabla \Theta)^2=\frac{{\bf p} \langle{\bf P}\rangle}{m^*_I}.
\end{eqnarray}
An integral in the l.h.s of this equation can be identically rewritten as follows
\begin{eqnarray}
&&\int_{r\le R} d{\bf r}n(\nabla \Theta)^2=\int_{r\le R} d{\bf r}n{\bf u}^2\nonumber\\
&&+\int_{r\le R} d{\bf r}n(\nabla [\Theta+{\bf ur}])\nabla [\Theta-{\bf ur}],
\end{eqnarray}
with $\bf u$ being arbitrary constant vector. The integrand in the last term organizes in the divergence of some vector field if we choose $\hbar{\bf u}={\bf p}m_r/m^*_I$ and then by using the Gauss theorem we obtain
\begin{eqnarray}
\int_{r\le R} d{\bf r}n{\bf u}^2
+\int_{r=R} d{\bf S}n(\Theta+{\bf ur})(\nabla\Theta-{\bf u})\nonumber\\
={\bf u}^2\frac{m_I}{m_r}\Delta_D,
\end{eqnarray}
where additional notation is used
\begin{eqnarray}\label{Delta_Df}
\Delta_D=\frac{m_r}{m_I}\left\{\int d{\bf r}(n-n_{\infty})-\Omega_DA_Dn_{\infty}\right\},
\end{eqnarray}
with $\Omega_D$ being the solid angle in $D$ dimensions. In 3D this formula was derived for a first time but in a different way by Gross \cite{Gross_62,note1} in context of a single ion immersed in superfluid $^4$He. Combining everything together we obtain, after some algebra, the average momentum of bosons
\begin{eqnarray}
\langle{\bf P}\rangle=\frac{{\bf p}\Delta_D}{1+\Delta_D},
\end{eqnarray}
valid up to leading order at small ${\bf p}$ and immediately the Bose polaron effective mass (\ref{m_eff})
\begin{eqnarray}\label{m*_I}
\frac{m^*_I}{m_I}=1+\Delta_D.
\end{eqnarray}
Importantly that parameters of the low-energy impurity spectrum, namely, the binding energy and the effective mass fully depend on the density profile of Bose particles with the immersed motionless impurity. The calculations of the effective mass additionally require the knowledge of the boundary-condition dependent constant $A_D$, which in turn, requires solution of Eq.~(\ref{eq_theta}) for a fixed density profile $n(r)$ of bosons.

From practical point of view it is more convenient to use another (nonetheless an equivalent) way of the effective mass computations. The formal solution of Eq.~(\ref{eq_Theta}) is a sum of general solution of homogeneous equation, $\Theta_0({\bf r})$, and partial solution of the non-homogeneous one, which is most easily constructed by the Green function method
\begin{eqnarray}\label{Theta}
\hbar\Theta({\bf r})=\hbar\Theta_0({\bf r})+\frac{m_r}{m^*_I}\int d{\bf r}'G({\bf r},{\bf r}'){\bf p}\nabla' n(r'),
\end{eqnarray}
where symmetric function $G({\bf r},{\bf r}')$ satisfies an equation
$\nabla\left\{ n(r) \nabla\right\} G({\bf r},{\bf r}')=\delta({\bf r}-{\bf r}')$. The later substitution in the l.h.s of Eq.~(\ref{identity}) with the following integration by parts and averaging over the directions of ${\bf p}$ yield the formal formula for the effective mass parameter
\begin{eqnarray}\label{Delta_op}
\Delta_D=\frac{m_r}{m_I}\frac{1}{D}\int d{\bf r}(\nabla n)\frac{1}{-\nabla n\nabla}(\nabla n).
\end{eqnarray}
Note that $\Theta_0({\bf r})$ fell out of final result. When the boson-impurity interaction is weak ($a^D_I\bar{n}\ll 1$), Bose particles are almost undisturbed and their local density slightly differs from the average one $\bar{n}$. It is believed then one can expand $n(r)=\bar{n}+\delta n(r)$ in the denominator of Eq.~(\ref{Delta_op}) and treat $\delta n(r)/\bar{n}$ as a `small' parameter. A few first terms of this series expansion preliminary converted in the Fourier space read
\begin{eqnarray}
\Delta_D=\frac{m_r}{m_I}\int d{\bf r}\frac{(\delta n)^2}{D\bar{n}}\left\{1-\frac{\delta n}{D\bar{n}}+\left(\frac{\delta n}{D\bar{n}}\right)^2\pm \ldots\right\}.
\end{eqnarray}
In such a way it could be explicitly shown for all higher-order terms with increasing powers of $\delta n/\bar{n}$ that the outlined tendency is preserved and we finally obtain
\begin{eqnarray}\label{Delta_D}
\Delta_D=\frac{m_r}{m_I}\int d{\bf r}\frac{(\delta n)^2}{D\bar{n}+\delta n}.
\end{eqnarray}
The latter expression is one of the central results of present study which in a case of $D=1$ reproduces recently published \cite{Pastukhov_3BIBP,Panochko_19} formula for the MF impurity effective mass in one dimension. Another interesting and analytically tractable example is the limit of almost incompressible Bose medium (for instance, liquid $^4$He), where 
\begin{eqnarray}
n(r)\simeq\left\{\begin{array}{c}
0, \ \ r\le a_I\\
\bar{n}, \ \  r>a_I
\end{array}\right.,
\end{eqnarray} 
and the appropriate effective mass
\begin{eqnarray}
\frac{m^*_I}{m_I}\simeq 1+\frac{m_r}{m_I}\frac{\Omega_Da^D_I\bar{n}}{(D-1)D},
\end{eqnarray} 
up to a factor $m_r/m_I$ in the second term coincides with the virtual mass of sphere moving in ideal classical fluid.

\section{Numerical results and discussion}

\subsection{3D system}
\begin{figure}[h!]
	\includegraphics[width=0.45\textwidth,clip,angle=-0]{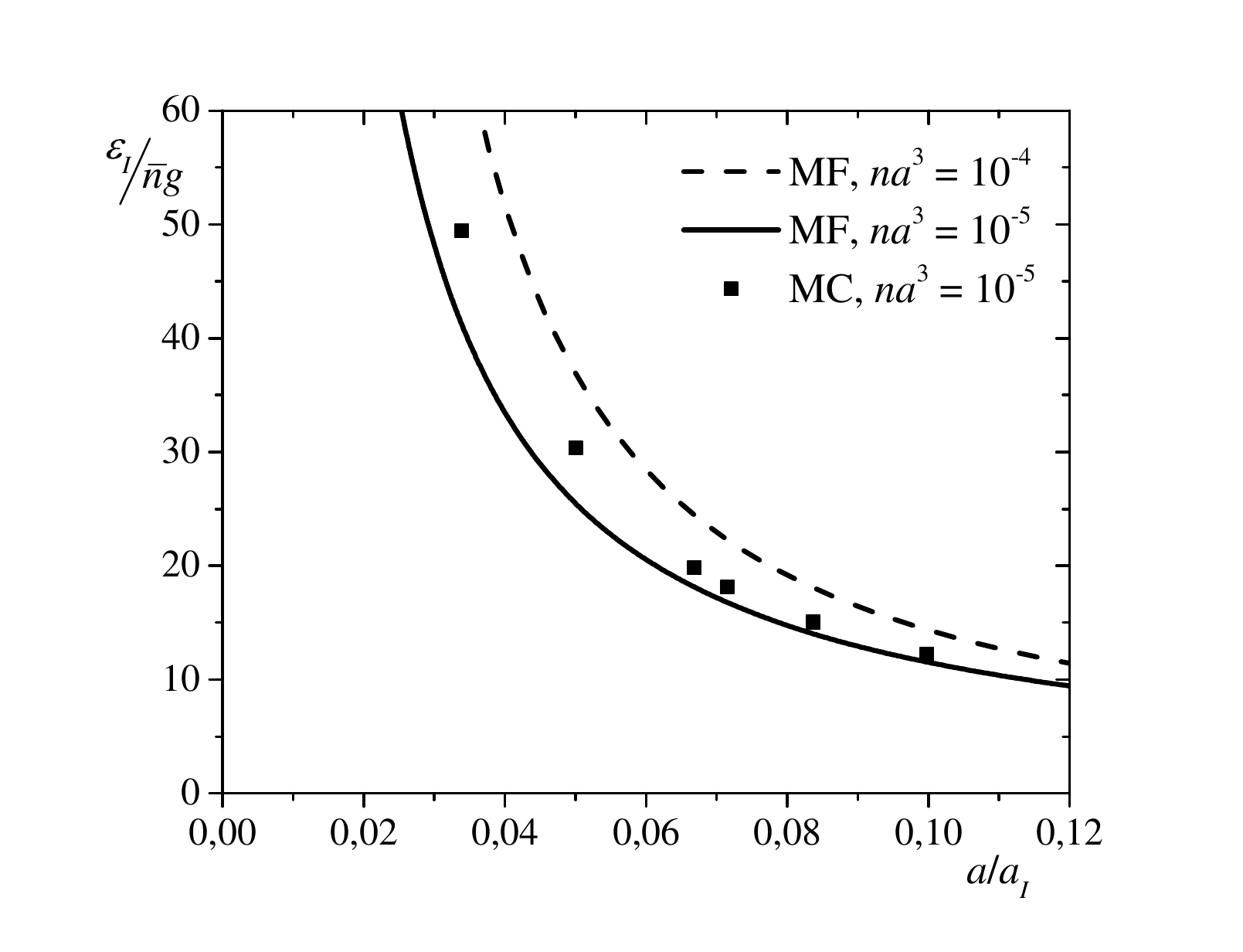}
	\caption{Binding energy of the repulsive 3D Bose polaron versus its radius $a_I/a$ for two values of dimensionless bosonic coupling $na^3=10^{-5}$ (solid line) and $na^3=10^{-4}$ (dashed line). Squares represent results of the MC simulations from \cite{Pena_Ardila_15} for the hard-sphere model.}
	\label{energy_all}
\end{figure}

\begin{figure}[h!]
	\includegraphics[width=0.45\textwidth,clip,angle=-0]{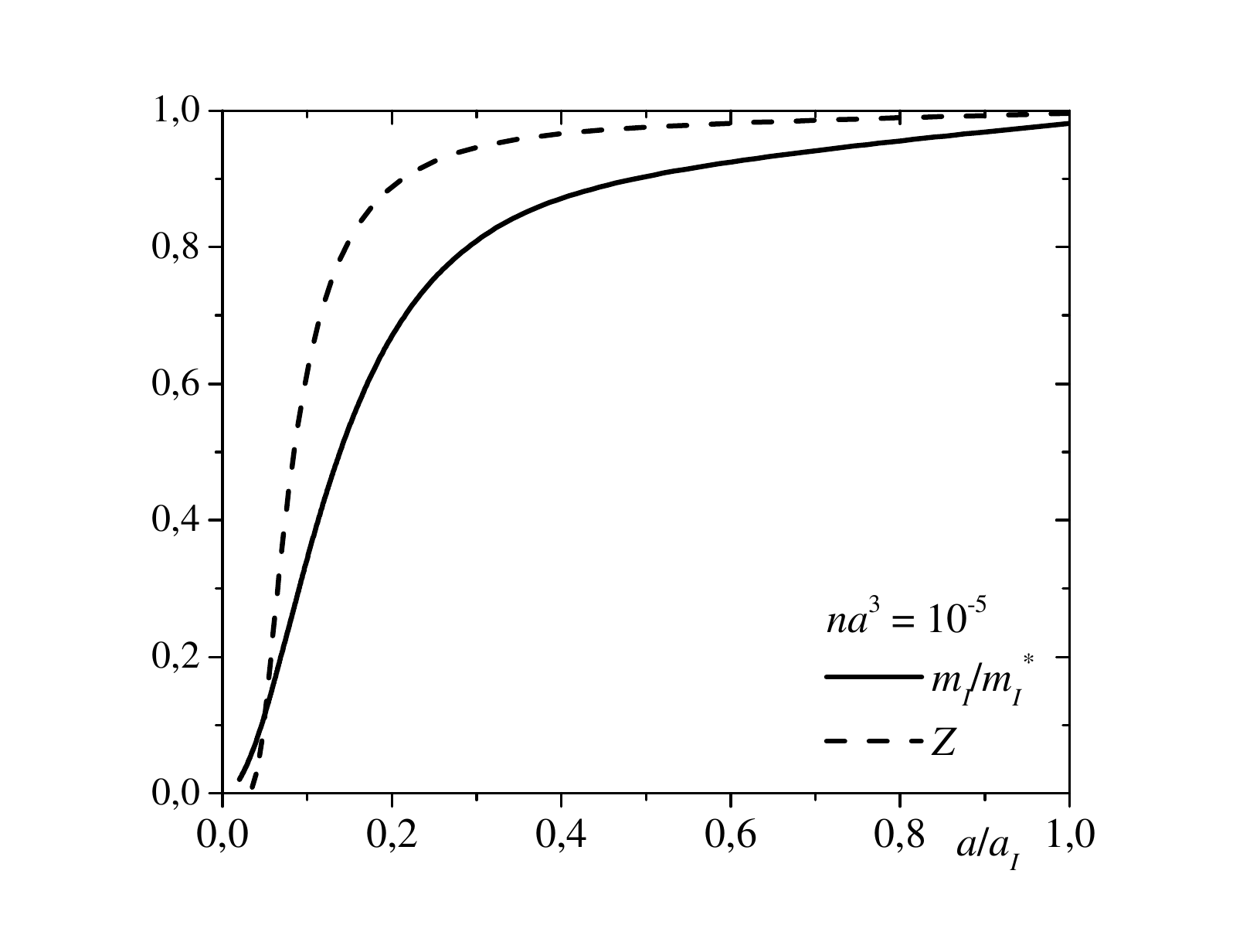}
	\caption{Inverse effective mass and quasiparticle residue of an impurity immersed in dilute 3D Bose gas, $na^3=10^{-5}$.}
	\label{mass_Z_10_5}
\end{figure}

\begin{figure}[h!]
	\includegraphics[width=0.45\textwidth,clip,angle=-0]{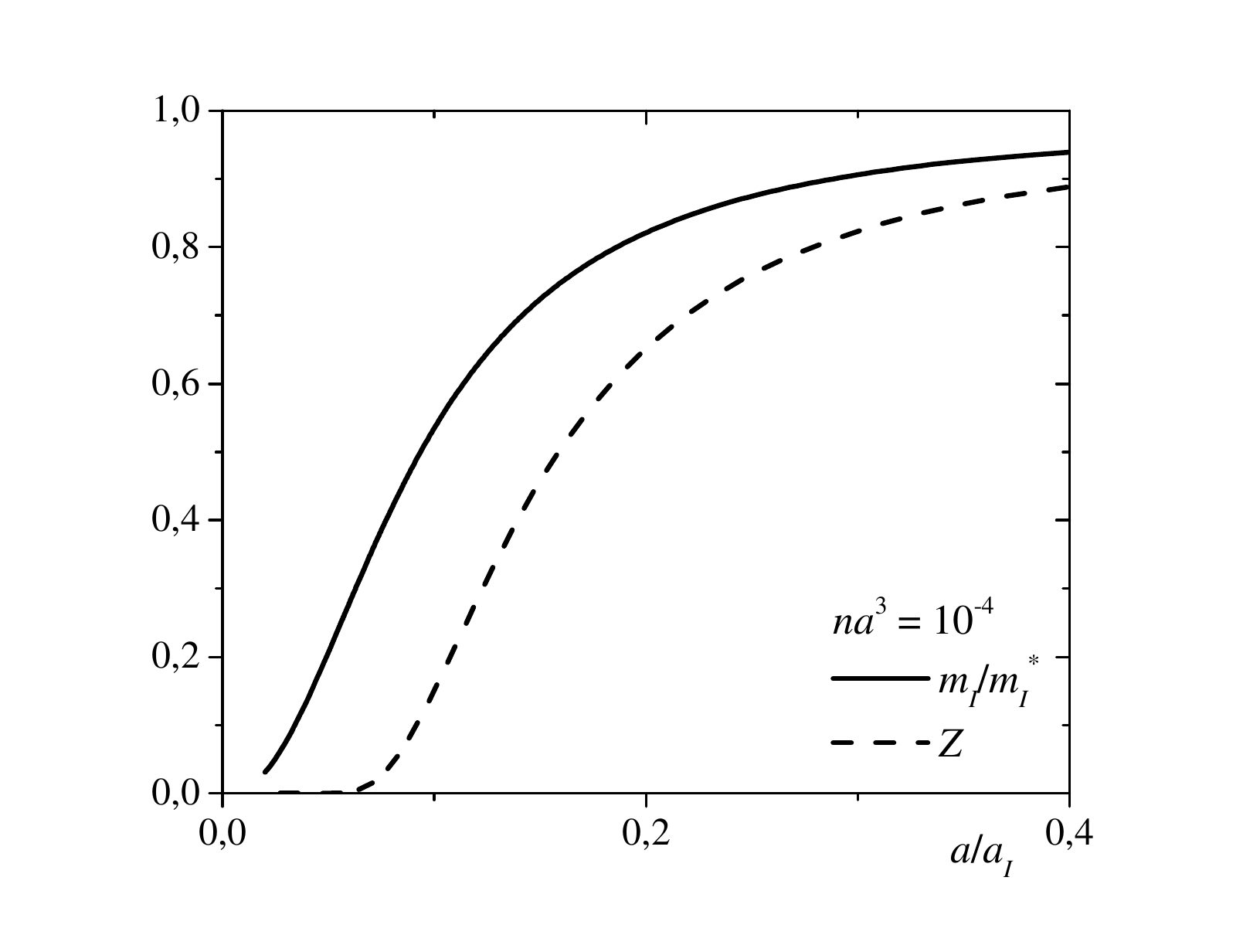}
	\caption{Inverse effective mass and quasiparticle residue of an impurity immersed in 3D system of bosons with gas parameter $na^3=10^{-4}$.}
	\label{mass_z_10_4}
\end{figure}

The 3D Bose polaron is the most well-studied both experimentally and theoretically system of an impurity immersed in an environment formed by weakly-\cite{Grusdt_Demler_15,Panochko_17} and strongly-interacting \cite{Panochko_18} bosons. Our MF treatment of this problem requires knowledge of the energy density of 3D Bose system alone. Although we focus on the dilute limit, where energy density is given by
\begin{eqnarray}\label{E_B3D}
\mathcal{E}[n]=\frac{2\pi\hbar^2a}{m}n^2,
\end{eqnarray}
(here $a$ is the $s$-wave scattering length) the presented formulation can be extended \cite{work_in_progress} to the so-called unitary limit, where formally $a\to \infty$ and $\mathcal{E}[n]\propto n^{5/3}$ in 3D. The next step, after identifying of $\mathcal{E}[n]$, is to calculate the ground-state wave function $\phi({\bf r})$, i.e., to solve Eq.~(\ref{phi}), which in dimensionless units looks as follows
\begin{eqnarray}\label{tilde_phi3D}
-\frac{1}{2}\frac{d^2\tilde{\phi}}{dx^2}-\frac{1}{x}\frac{d\tilde{\phi}}{dx}+{\tilde{\phi}}^3=\tilde{\phi},
\end{eqnarray}
where $\phi({\bf r})=\sqrt{n_{\infty}}\tilde{\phi}(\kappa r)$, $x=\kappa r$, $\kappa^2=m_r\mu/\hbar^2$, $\mu=4\pi\hbar^2n_{\infty}/m$ and we only seek for the spherically symmetric solutions.
This equation should be supplemented by the boundary conditions $\tilde{\phi}(\kappa a_I)=0$ and $\tilde{\phi}(\infty)=1$, while the density of bosons at infinity $n_{\infty}$ and the chemical potential should be related to average density $\bar{n}L^3=n_{\infty}\int d{\bf r}\tilde{\phi}^2(\kappa r)$. For the numerical calculations we chose the following set of parameters \cite{Pena_Ardila_15}: $na^3=10^{-5}$, $m_I=m$. And varying the radius of hard-sphere potential $a_I$ in units of bosonic scattering length $a$ we have obtained the appropriate density profiles $|\phi({\bf r})|^2$ that allowed us to calculate parameters of the impurity spectrum $\varepsilon_I$ (\ref{E}) and $m^*_I$ (\ref{m*_I}), (\ref{Delta_D}) as well as quasiparticle residue $Z$ (\ref{Z}). The results are presented in Figs.~\ref{energy_all}, \ref{mass_Z_10_5}, \ref{mass_z_10_4}.
Particularly, in Fig.~\ref{energy_all} black squares denote the Bose polaron binding energy obtained in MC simulations \cite{Pena_Ardila_15}. It is worth mentioning that the effective mass presented in \cite{Pena_Ardila_15} was calculated for the model with square-well boson-impurity potential, therefore not presented here \cite{note2}. For comparison, we have also plotted parameters of the impurity spectrum when the interaction between bosons is a bit stronger, $na^3=10^{-4}$. In this case the MF calculations are also believed to be valid because the beyond-mean-field correction to the chemical potential of the uniform Bose system is only of order $6\%$ magnitude.

\subsection{2D system}
\begin{figure}[h!]
	\includegraphics[width=0.45\textwidth,clip,angle=-0]{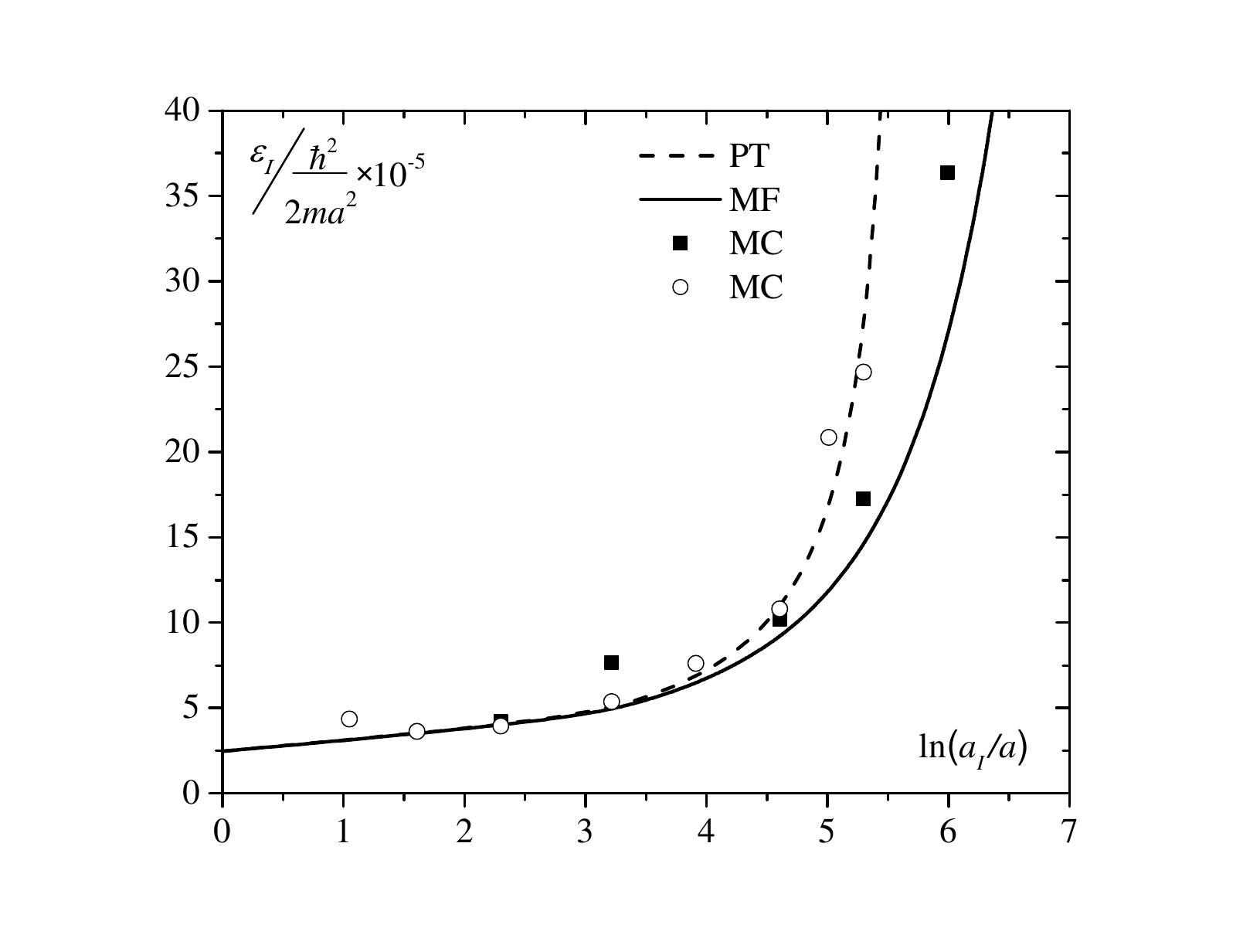}
	\caption{Binding energy of the 2D Bose polaron as a function of $a_I/a$ at fixed dimensionless bosonic coupling $na^2=10^{-5}$ (solid line). Results of the MC simulations \cite{Akaturk} for a model with the short range boson-impurity repulsion and two types of trial wave functions are denoted by circles and squares, respectively. Dashed line represent the first-order perturbation theory (PT) calculations \cite{Pastukhov_2DBP}.}
	\label{E_2D}
\end{figure}

\begin{figure}[h!]
	\includegraphics[width=0.45\textwidth,clip,angle=-0]{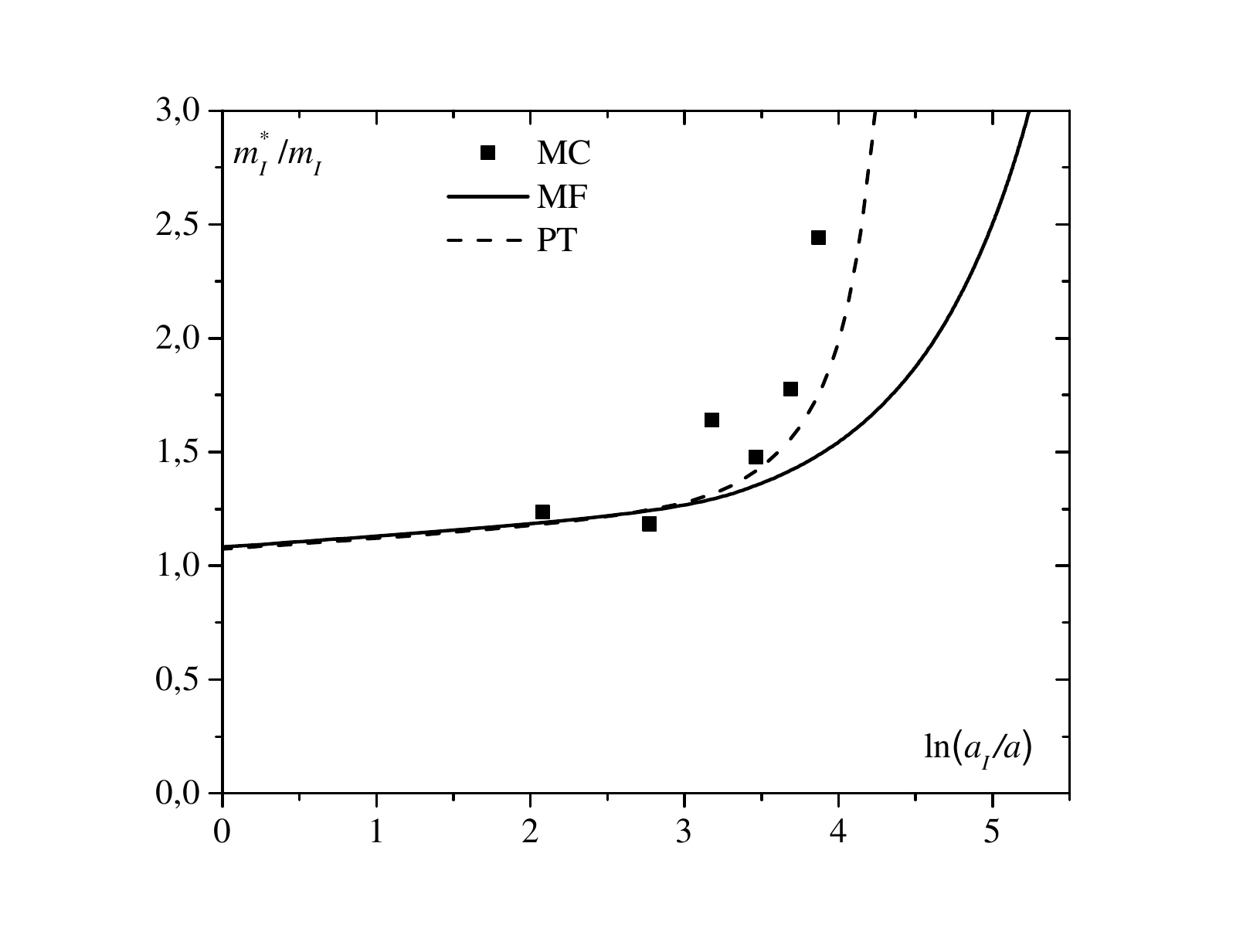}
	\caption{The 2D Bose polaron inverse effective mass at dimensionless boson-boson coupling $na^2=10^{-5}$. Squares stand for the MC data and are taken from \cite{Akaturk}. Solid and dashed lines are the results of present MF calculations Eq.~(\ref{Delta_D}) and the first-order perturbation theory \cite{Pastukhov_2DBP}, respectively.}
	\label{mass_2D}
\end{figure}

\begin{figure}[h!]
	\includegraphics[width=0.45\textwidth,clip,angle=-0]{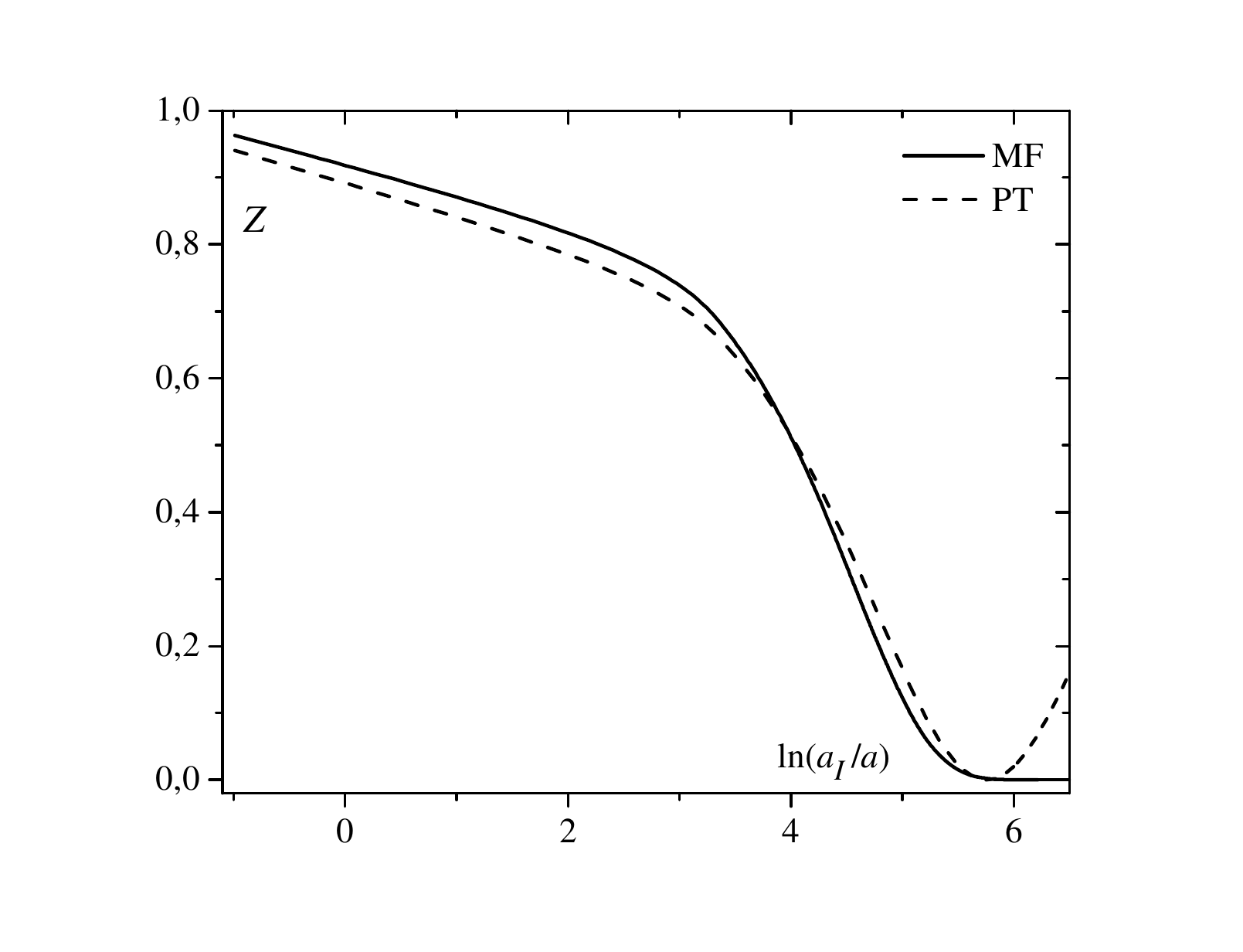}
	\caption{The quasiparticle residue of an impurity immersed in 2D bosonic system with gas parameter $na^2=10^{-5}$. Solid and dashed lines display the present calculations and the perturbative result \cite{Pastukhov_2DBP}, respectively.}
	\label{Z_2D}
\end{figure}

The MF energy density of 2D Bose gas which is valid only in the extremely dilute limit $na^2\ll 1$ reads \cite{Schick,Lozovik,Astrakharchik_09,Pastukhov_19}
\begin{eqnarray}\label{E_B2D}
\mathcal{E}[n]=\frac{2\pi\hbar^2n^2/m}{|\ln na^2|},
\end{eqnarray}
where $s$-wave scattering length $a$ characterizes short-range interaction between particles in 2D. Substituting this expression in Eq.~(\ref{phi}) and neglecting subleading terms of order $1/|\ln na^2|^2$ in $\mathcal{E}'[n]$ we obtain the second-order nonlinear differential equation [here $x$ and $\kappa$ are the same as in 3D case, but $\mu=4\pi\hbar^2n_{\infty}/(m|\ln n_{\infty}a^2|)$]
\begin{eqnarray}\label{tilde_phi2D}
-\frac{1}{2}\frac{d^2\tilde{\phi}}{dx^2}-\frac{1}{2x}\frac{d\tilde{\phi}}{dx}+\frac{\tilde{\phi}^3}{1+|\ln\tilde{\phi}^2|/|\ln n_{\infty}a^2|}=\tilde{\phi}
\end{eqnarray}
that should be solved with the same as Eq.~(\ref{tilde_phi3D}) boundary conditions. Again, with the bosonic density profiles in hands we numerically computed low-momentum characteristics of the 2D Bose polaron (see Figs.~\ref{E_2D},\ref{mass_2D},\ref{Z_2D}).
In order to make connection to results of recent MC simulations \cite{Akaturk} of impurity in 2D system of bosons we took the same set of parameters, characterizing the system under consideration, namely, $na^2=10^{-5}$ and $m_I=m$. This is a very important test for our MF approach because in their MC study \cite{Akaturk} the authors used hard-disk potential for modeling both the boson-boson and the boson-impurity two-body interactions. Of course, the magnitude of dimensionless inter-boson coupling $na^2=10^{-5}$ is too large to describe bosons accurately by the MF energy density (\ref{E_B2D}) and one has to take into account the beyond-MF corrections \cite{Astrakharchik_09,Pastukhov_19} in Eq.~(\ref{tilde_phi2D}). It is understood, however, that the qualitative MF picture of the impurity behavior will not be changed even in the latter case. The MC calculations of the binding energy of 2D Bose polaron were performed in \cite{Akaturk} for two types of wave functions and therefore two types of symbols are plotted in Fig.~\ref{E_2D}. We also supplied figures with curves (dashed lines) representing the perturbation theory calculations \cite{Pastukhov_2DBP} valid for extremely dilute Bose systems and weak boson-impurity repulsion. The latter restriction is crucial, because values $\ln(a_I/a)=5\div 6$ are limiting ones for the perturbative results. Particularly, starting from this region the effective mass and the quasiparticle residue behave unphysically. In contrast, our MF calculations are free of these shortcomings and can be applied for description (at least qualitative) of Bose polarons at any magnitude of fraction $a_I/a$.

\begin{figure}[h!]
	\includegraphics[width=0.45\textwidth,clip,angle=-0]{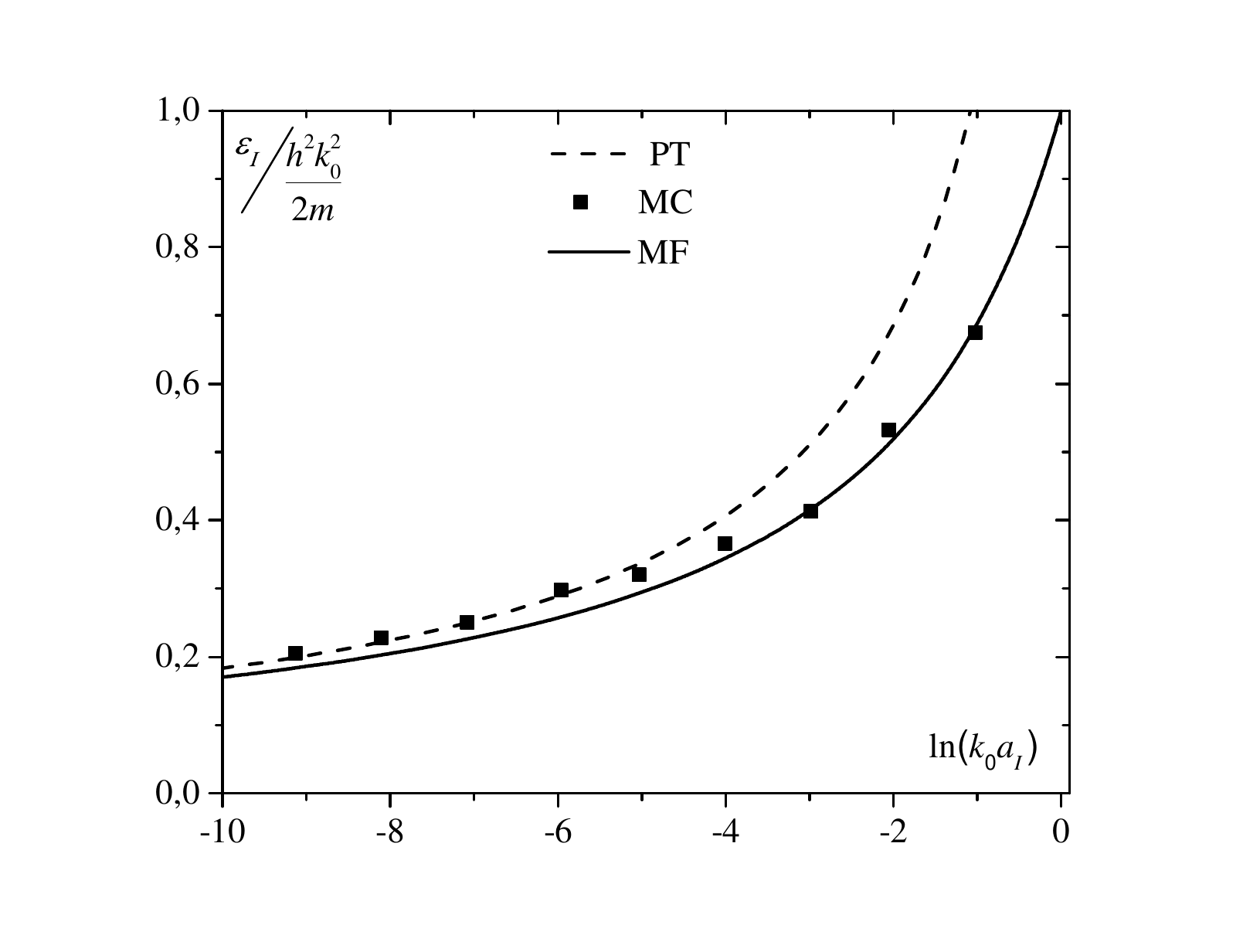}
	\caption{Energy of an impurity immersed in the dilute 2D Bose gas ($na^2=10^{-40}$) versus dimensionless parameter $\ln(k_0a_I)$ (solid line). Squares stand for the MC results from Ref.~\cite{Pena_Ardila_Astrakharchik_19}, while dashed line represent the first-order perturbation theory calculations \cite{Pastukhov_2DBP}.}
	\label{en}
\end{figure}

\begin{figure}[h!]
	\includegraphics[width=0.45\textwidth,clip,angle=-0]{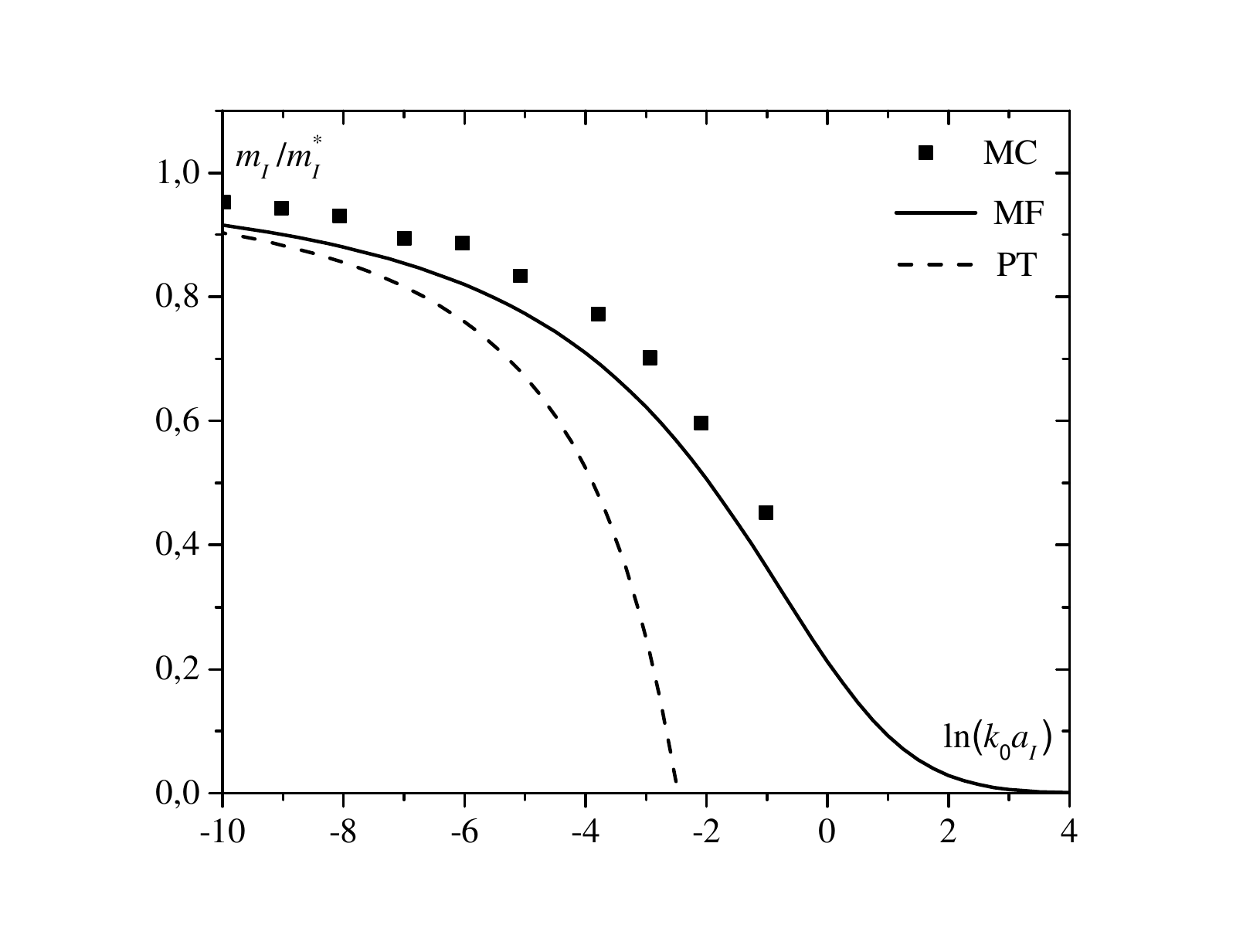}
	\caption{Impurity effective mass at dimensionless boson-boson coupling $na^2=10^{-10}$. Notations are similar to those in Fig.~\ref{en}.}
	\label{mm}
\end{figure}

\begin{figure}[h!]
	\includegraphics[width=0.45\textwidth,clip,angle=-0]{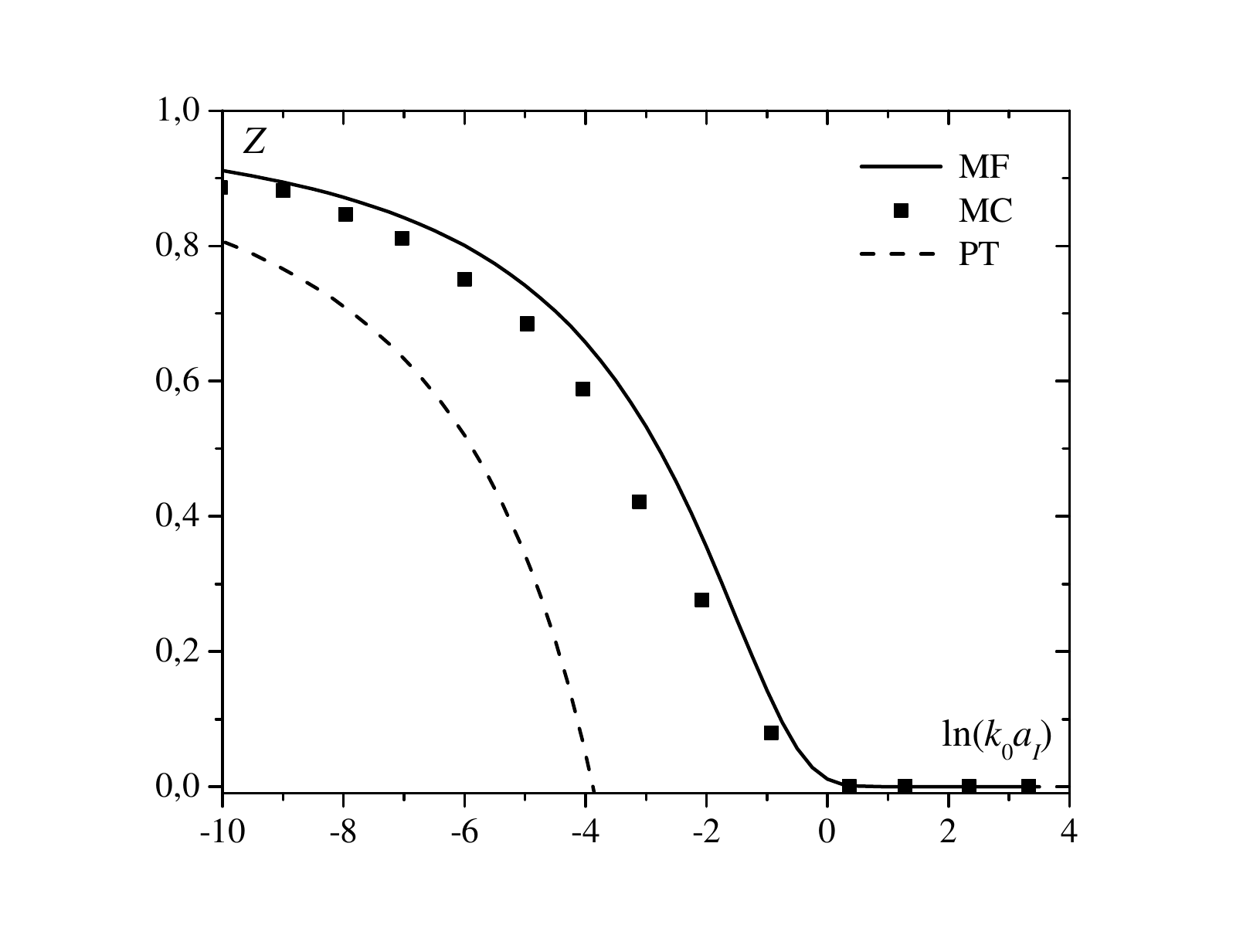}
	\caption{The quasiparticle residue of the 2D Bose polaron at $na^2=10^{-40}$. Squares, and dashed line stand for the MC data \cite{Pena_Ardila_Astrakharchik_19} and the perturbation theory results \cite{Pastukhov_2DBP}, respectively.}
	\label{z}
\end{figure}

One typically believes that the MF approximation should be working better for an extremely dilute bosonic environments. It is instructive, therefore, to test our results by comparison with the MC simulations of the 2D Bose polaron performed in Ref.~\cite{Pena_Ardila_Astrakharchik_19}. The dimensionless boson-boson coupling parameter was chosen there to be $na^2=10^{-40}$, while the boson-impurity interaction was taken into account by the 2D analogue of the Bethe-Peierls boundary condition. This pseudo-potential approximately models the two-body spherical-well potential, and at small $s$-wave scattering lengths can be effectively used instead of a hard-sphere interaction. But it is not well-suited for our formulation, where the non-linear terms in Eqs.~(\ref{tilde_phi3D}), (\ref{tilde_phi2D}) provide strong singularities at the origin. Numerical results of the MF calculations together with the first-order perturbative curves \cite{Pastukhov_2DBP} for the parameters of the Bose polaron spectrum and for its residue are presented in Figs.~\ref{en},\ref{mm},\ref{z}. 
Here we again assumed the equal mass $m_I=m$ limit and introduced the auxiliary scale $k_0=\sqrt{2\pi n}$. Although, the hard-sphere interaction is not equivalent to the pseudo-potential provided by the Bethe-Peierls boundary condition and particularly cannot describe the system at large positive $\ln(k_0a_I)$, we see that the MF approximation gives a better agreement with MC results than the standard perturbative theory especially when the boson-impurity interaction increases.

The most unexpected consequence of the whole previous analysis is that in general the MF approximation in context of the impurity immersed in 2D and 3D Bose gases demonstrates worse consistence with MC results in comparison with its application to the 1D Bose polaron problem \cite{Panochko_19}. This might look particularly strange because there is always a thought that the MF should work better in higher dimensions. So, what is wrong with MF in this case? To answer this question we first would like to stress that physics of Bose gas itself is much more complicated in higher dimensions than in 1D. Even when the system is dilute enough the echoes of many-body effects are tangible in the thermodynamics of bosons. In order to argue the latter statement let us recall that the perturbative result for the ground-state energy of 1D Bose gas has much more vast region of applicability than its high-dimensional counterparts. This suggests that few-body quantum effects in higher dimensions which are typically missed in the MF treatment should also affect the polaron properties. In contrast, in 1D the first quantum corrections to the impurity spectrum were recently found \cite{Jager} to be relatively small. The possible way for an inclusion of the many-boson effects is to extend the presented MF treatment to the so-called local density approximation. Technically this approximation is nothing but the replacement of the MF energy density of bosons in Hamiltonian (\ref{H_B}) by more sophisticated formula, which necessarily contains corrections due to quantum fluctuations.

\section{Concluding remarks}
In summary, we have proven by calculating parameters of the low-momentum spectrum of a single impurity immersed in dilute two- and three-dimensional Bose condensates that the mean field approximation is a promising semi-quantitative tool for analysis of various aspects of the Bose polaron problem even in the strong-coupling regime. The main advantage of this approach is its simplicity which in combination with very clear physical interpretation give hope for further application to the impurity problem in strongly-interacting Bose condensates and fermionic superfluids. Our study also provides the indirect evidence of importance of the quantum fluctuations for the accurate quantitative description of Bose polarons in two and three dimensions. Having analyzed parameters of the impurity spectrum we can make conclusions about a general tendency of the mean field results. Particularly, comparing the calculated binding energy and the effective mass with the results of Monte Carlo simulations both in two and three dimensions, we have found out that the mean field approximation always provides the lower bond for these two parameters, while slightly overestimating a magnitude of the quasiparticle residue.

\begin{center}
	{\bf Acknowledgements}
\end{center}
We are indebted to Prof.~B.~Tanatar, Dr.~E.~Akaturk, Dr.~L.~A.~Pe\~na Ardila and Prof.~G.~E.~Astrakharchik for providing us with their results of Monte Carlo simulations. Work of O.~H. was partly supported by Project FF-83F (No. 0119U002203) from the Ministry of Education and Science of Ukraine.


\begin{thebibliography}{99}
	
\bibitem{Gross_61} E.~P.~Gross, 
\href{https://doi.org/10.1007/BF02731494}{IL Nuovo Cimento {\bf 20}, 454 (1961).}
\bibitem{Pitaevskii_61} L.~P.~Pitaevskii, 
\href{http://www.jetp.ac.ru/cgi-bin/dn/e_013_02_0451.pdf}{Sov. Phys. JETP {\bf 13}, 451 (1961).}
\bibitem{Dalfovo_etal} F.~Dalfovo, S.~Giorgini, L.~P. Pitaevskii, and S.~Stringari,
\href{https://doi.org/10.1103/RevModPhys.71.463}{Rev.~Mod.~Phys. {\bf 71}, 463 (1999).}

\bibitem{Astrakharchik_04} G.~E.~Astrakharchik and L.~P.~Pitaevskii, 
\href{https://doi.org/10.1103/PhysRevA.70.013608}{Phys.~Rev.~A {\bf 70}, 013608 (2004).}


\bibitem{Cucchietti_06} F.~M.~Cucchietti and E.~Timmermans,
\href{https://doi.org/10.1103/PhysRevLett.96.210401}{Phys.~Rev.~Lett. {\bf 96}, 210401 (2006).}

\bibitem{Kalas_06} R.~M.~Kalas and D.~Blume, 
\href{https://doi.org/10.7566/JPSJ.87.043002}{Phys.~Rev.~A {\bf 73}, 043608 (2006).}

\bibitem{Sacha_06} K.~Sacha and E.~Timmermans, 
\href{https://doi.org/10.1103/PhysRevA.73.063604}{Phys.~Rev.~A {\bf 73}, 063604 (2006).}

\bibitem{Bruderer_08} M.~Bruderer, W.~Bao, D.~Jaksch, 
\href{https://doi.org/10.1209/0295-5075/82/30004}{EPL (Europhysics Letters) {\bf 82}, 30004 (2008).}

\bibitem{Roberts_09} D.~C.~Roberts and S.~Rica, 
\href{https://doi.org/10.1103/PhysRevLett.102.025301}{Phys.~Rev.~Lett. {\bf 102}, 025301 (2009).}

\bibitem{Blinova_13} A.~A.~Blinova, M.~G.~Boshier, and E.~Timmermans, 
\href{https://doi.org/10.1103/PhysRevA.88.053610}{Phys.~Rev.~A {\bf 88}, 053610 (2013).}

\bibitem{Gross} E.~P.~Gross,
\href{https://doi.org/10.1016/0003-4916(58)90037-X}{ Ann.~Phys. {\bf 4}, 57 (1958)}
; \href{https://doi.org/10.1063/1.1703944}{J.~Math.~Phys. {\bf 4}, 195 (1963).}



\bibitem{Volosniev_17} A.~G.~Volosniev, H.-W.~Hammer, 
\href{https://doi.org/10.1103/PhysRevA.96.031601}{Phys.~Rev.~A {\bf 96},
	031601(R) (2017).}

\bibitem{Pastukhov_3BIBP} V.~Pastukhov, 
\href{https://doi.org/10.1016/j.physleta.2019.05.018}{Phys.~Lett.~A {\bf 383}, 2610 (2019).}

\bibitem{Smith} D.~H.~Smith and A.~G.~Volosniev, 
\href{https://doi.org/10.1103/PhysRevA.100.033604}{Phys. Rev. A {\bf 100}, 033604 (2019).}
\bibitem{Jorgensen} N.~B.~Jorgensen, L.~Wacker, K.~T.~Skalmstang, M.~M.~Parish, J.~Levinsen, R.~S.~Christensen, G.~M.~Bruun and J.~J.~Arlt,
\href{https://doi.org/10.1103/PhysRevLett.117.055302}{Phys.~Rev.~Lett.  {\bf 117}, 055302 (2016).}

\bibitem{Hu} M.-G.~Hu, M.~J.~Van de Graaff, D.~Kedar, J.~P.~Corson, E.~A.~Cornell and D.~S.~Jin, 
\href{https://journals.aps.org/prl/abstract/10.1103/PhysRevLett.117.055301}{Phys.~Rev.~Lett. {\bf 117}, 055301 (2016).}

\bibitem{Novikov_09} A.~Novikov and M.~Ovchinnikov, 
\href{https://doi.org/10.1088/1751-8113/42/13/135301}{J.~Phys.~A: Math.~Theor. {\bf 42}, 135301 (2009).}

\bibitem{Novikov_10} A.~Novikov and M.~Ovchinnikov,
\href{https://doi.org/10.1088/0953-4075/43/10/105301}{ J.~Phys.~B: At.~Mol.~Opt.~Phys. {\bf 43}, 105301 (2010).}
\bibitem{Rath_13} S.~P.~Rath and R.~Schmidt, 
\href{https://doi.org/10.1103/PhysRevA.88.053632}{Phys.~Rev.~A {\bf 88}, 053632 (2013).}

\bibitem{Shashi} A.~Shashi, F.~Grusdt, D.~A. Abanin, and E.~Demler,
\href{https://doi.org/10.1103/PhysRevA.89.053617}{Phys.~Rev.~A {\bf 89}, 053617 (2014).}
\bibitem{Li_14} W.~Li and S.~Das~Sarma, 
\href{https://doi.org/10.1103/PhysRevA.90.013618}{Phys.~Rev.~A {\bf 90}, 013618 (2014).}

\bibitem{Christensen_15} R.~S.~Christensen, J.~Levinsen, and G.~M.~Bruun,
\href{https://doi.org/10.1103/PhysRevLett.115.160401}{ Phys.~Rev.~Lett. {\bf 115}, 160401 (2015).}

\bibitem{Grusdt_15} F. Grusdt, Y. E. Shchadilova, A. N. Rubtsov, and
E. Demler,  
\href{https://www.nature.com/articles/srep12124}{Sci.~Rep. {\bf 5}, 12124 (2015).}

\bibitem{Vlietinck_15} J. Vlietinck W.~Casteels, K.~Van~Houcke, J.~Tempere, J.~Ryckebusch, and J.~T.~Devreese, 
\href{https://doi.org/10.1088/1367-2630/17/3/033023}{New~J.~Phys. {\bf 17},  033023 (2015).}

\bibitem{Pena_Ardila_15} L.~A.~Pe\~na Ardila and S.~Giorgini, 
\href{https://doi.org/10.1103/PhysRevA.92.033612}{Phys.~Rev.~A {\bf 92}, 033612 (2015).}

\bibitem{Shchadilova} Y.~E.~Shchadilova, F.~Grusdt, A.~N.~Rubtsov, and E.~Demler
\href{https://doi.org/10.1103/PhysRevA.93.043606}{Phys.~Rev.~A {\bf 93}, 043606 (2016).}

\bibitem{Pena_Ardila_16} L.~A.~Pe\~na Ardila and S.~Giorgini, 
\href{https://doi.org/10.1103/PhysRevA.94.063640}{Phys.~Rev.~A {\bf 94}, 063640 (2016).}

\bibitem{GSSD} F.~Grusdt, R.~Schmidt, Y.~E.~Shchadilova, and E.~Demler, 
\href{https://doi.org/10.1103/PhysRevA.96.013607}{Phys.~Rev.~A {\bf 96}, 013607 (2017).}

\bibitem{Pena_Ardila_19} L.~A.~Pe\~na Ardila, N.~B.~J\o{}rgensen, T.~Pohl, S.~Giorgini, G.~M.~Bruun, and J.~J.~Arlt, 
\href{https://doi.org/10.1103/PhysRevA.99.063607}{Phys.~Rev.~A {\bf 99}, 063607 (2019).}


\bibitem{Levinsen_17} J.~Levinsen, M.~M.~Parish, R.~S.~Christensen, J.~J.~Arlt, and G.~M.~Bruun \href{https://doi.org/10.1103/PhysRevA.96.063622}{, Phys.~Rev.~A. {\bf 96}, 063622 (2017).}
\bibitem{Guenther} N.-E.~Guenther, P.~Massignan, M.~Lewenstein, G.~M.~Bruun, 
\href{https://doi.org/10.1103/PhysRevLett.120.050405}{Phys.~Rev.~Lett. {\bf 120}, 050405 (2018).}
\bibitem{Bosepolaron_D} V.~Pastukhov, \href{https://doi.org/10.1088/1751-8121/aab9c1}{J.~Phys.~A:~Math.~Theor. {\bf 51}, 195003 (2018).}
\bibitem{Field} B.~Field, J.~Levinsen, M.~M.~Parish, \href{https://doi.org/10.1103/PhysRevA.101.013623}{Phys. Rev. A {\bf 101}, 013623 (2020).}


\bibitem{Volosniev_15} A.~G.~Volosniev, H.-W.~Hammer, and N.~T.~Zinner, 
\href{https://doi.org/10.1103/PhysRevA.92.023623}{Phys.~Rev.~A {\bf 92}, 023623 (2015).}

\bibitem{Shchadilova_16} Y.~E.~Shchadilova, R.~Schmidt, F.~Grusdt, and E.~Demler, 
\href{https://doi.org/10.1103/PhysRevLett.117.113002}{Phys.~Rev.~Lett. {\bf 117}, 113002 (2016).}

\bibitem{Lampo_18} A.~Lampo, S.~H.~Lim, M.~\'A.~Garc\'ia-March, M.~Lewenstein, 
\href{https://doi.org/10.22331/q-2017-09-27-30}{Quantum {\bf 1}, 30 (2018).}

\bibitem{Lausch} T.~Lausch, A.~Widera, and M.~Fleischhauer,
\href{https://doi.org/10.1103/PhysRevA.97.023621}{Phys. Rev. A {\bf 97}, 023621 (2018).}

\bibitem{Drescher} M.~Drescher, M.~Salmhofer, and T.~Enss,
\href{https://doi.org/10.1103/PhysRevA.99.023601}{Phys.~Rev.~A {\bf 99}, 023601 (2019).}
\bibitem{Liu} W.~E.~Liu, J.~Levinsen, and M.~M.~Parish, 
\href{https://doi.org/10.1103/PhysRevLett.122.205301}{Phys. Rev. Lett. {\bf 122}, 205301 (2019).}
\bibitem{Boyanovsky} D.~Boyanovsky, D.~Jasnow, X.-L.~Wu, and R.~C.~Coalson
\href{https://doi.org/10.1103/PhysRevA.100.043617}{Phys. Rev. A {\bf 100}, 043617 (2019).}


\bibitem{Zinner} N.~T.~Zinner, EPL (Europhysics Letters) {\bf 101}, 60009 (2013). \href{https://doi.org/10.1209/0295-5075/101/60009}{ EPL (Europhysics Letters) {\bf 101}, 60009 (2013).}
\bibitem{Levinsen_15} J.~Levinsen, M.~M.~Parish, and G.~M.~Bruun, 
\href{https://doi.org/10.1103/PhysRevLett.115.125302}
{Phys. Rev. Lett. {\bf 115}, 125302 (2015). }
\bibitem{Sun} M.~Sun, H.~Zhai, and X.~Cui,
\href{https://doi.org/10.1103/PhysRevLett.119.013401}
{ Phys. Rev. Lett. {\bf 119}, 013401 (2017).}
\bibitem{Yoshida} S.~M.~Yoshida, S.~Endo, J.~Levinsen, M.~M.~Parish, 
\href{https://doi.org/10.1103/PhysRevX.8.011024}{Phys.~Rev.~X {\bf 8}, 011024 (2018).}
\bibitem{Naidon} P.~Naidon,  \href{https://doi.org/10.7566/JPSJ.87.043002}{J.~Phys.~Soc.~Jpn. {\bf 87} 043002 (2018).}



\bibitem{Pastukhov_2DBP} V.~Pastukhov,
\href{https://doi.org/10.1088/1361-6455/aacdcb}{ J.~Phys.~B:~At.~Mol.~Opt.~Phys. {\bf 51}, 155203 (2018).}
\bibitem{Grusdt_16} F.~Grusdt, 
\href{https://doi.org/10.1103/PhysRevB.93.144302}{Phys.~Rev.~B {\bf 93}, 144302 (2016).}
\bibitem{Grusdt_Fleischhauer_16} F.~Grusdt and M.~Fleischhauer, 
\href{https://doi.org/10.1103/PhysRevLett.116.053602}{Phys.~Rev.~Lett. {\bf 116}, 053602 (2016).}
\bibitem{Pena_Ardila_Pohl} L.~A.~Pe\~na Ardila and T.~Pohl, 
\href{hhttps://doi.org/10.1088/1361-6455/aaf35e}{J.~Phys.~B: At. Mol. Opt. Phys. {\bf 52}, 015004 (2019).}
\bibitem{Akaturk} E.~Akaturk and B.~Tanatar,
\href{https://doi.org/10.1142/S0217979219502382}{ Int.~J.~Mod.~Phys.~B {\bf 33}, 1950238 (2019).}
\bibitem{Pena_Ardila_Astrakharchik_19} L.~A.~Pe\~na Ardila, G.~E.~Astrakharchik, S.~Giorgini,
\href{https://doi.org/10.1103/PhysRevResearch.2.023405}{Phys. Rev. Research {\bf 2}, 023405 (2020).}




\bibitem{Panochko_19} G.~Panochko, V.~Pastukhov, 
\href{https://doi.org/10.1016/j.aop.2019.167933}{Ann.~Phys. {\bf 409}, 167933 (2019).}
\bibitem{Parisi_17} L.~Parisi and S.~Giorgini, 
\href{https://journals.aps.org/pra/abstract/10.1103/PhysRevA.95.023619}{Phys.~Rev.~A {\bf 95}, 023619 (2017).}
\bibitem{Grusdt_17} F.~Grusdt, G.~E.~Astrakharchik, E.~A.~Demler, 
\href{https://iopscience.iop.org/article/10.1088/1367-2630/aa8a2e/meta}{New~J.~Phys. {\bf 19}, 103035 (2017).}

\bibitem{Grusdt_Demler_15} F.~Grusdt and E.~A.~Demler, 
\href{https://arxiv.org/abs/1510.04934}{{\it Proceedings of the International School of Physics `Enrico Fermi'}, arXiv:1510.04934.}
\bibitem{Panochko_17} G.~Panochko, V.~Pastukhov, I.~Vakarchuk,
\href{http://www.icmp.lviv.ua/journal/zbirnyk.89/13604/abstract.html}{ Condens.~Matter~Phys. {\bf  20}, 13604 (2017).}

\bibitem{Landau} L.~D.~Landau, J. Phys. USSR {\bf 5}, 71 (1941).

\bibitem{LLP} T.~D.~Lee, F.~E.~Low, and D.~Pines,
\href{https://doi.org/10.1103/PhysRev.90.297}{Phys. Rev. {\bf 90}, 297 (1953).}

\bibitem{Pitaevskii_Stringari} L.~P.~Pitaevskii and S.~Stringari, {\it Bose-Einstein Condensation} (Oxford University Press, 2003).

\bibitem{Braz} J.~E.~H.~Braz, H.~Tercas,
\href{https://doi.org/10.1103/PhysRevA.101.023607}{Phys. Rev. A {\bf 101}, 023607 (2020).}

\bibitem{Panochko_Pastukhov} G.~Panochko and V.~Pastukhov,\href{https://arxiv.org/abs/1909.01256}{arXiv:1909.01256.}  

\bibitem{Boudjemaa} A. Boudjemaa, 
\href{https://doi.org/10.1103/PhysRevA.90.013628}{Phys. Rev. A {\bf 90}, 013628 (2014).}

\bibitem{Takahashi} J. Takahashi, R. Imai, E. Nakano, and K. Iida,
\href{https://doi.org/10.1103/PhysRevA.100.023624}{Phys. Rev. A {\bf 100}, 023624 (2019).}




\bibitem{Hydrodynamcs} L.~D.~Landau, E.~M.~Lifshitz, {\it Fluid Mechanics} (Pergamon Press, New York, 1987).	





\bibitem{Gross_62} E.~P.~Gross, 
\href{https://doi.org/10.1016/0003-4916(62)90217-8}{Ann.~Phys. {\bf 19}, 234 (1962).}
\bibitem{note1} We have learned about work of Gross only during final completion of the manuscript.



\bibitem{Panochko_18} G.~Panochko, V.~Pastukhov, I.~Vakarchuk, 
\href{https://www.worldscientific.com/doi/abs/10.1142/S0217979218500534}{Int. J. Mod. Phys. B {\bf 32}, 1850053 (2018).}


\bibitem{work_in_progress} O.~Hryhorchak, G.~Panochko and V.~Pastukhov,
\href{https://arxiv.org/abs/2006.06188}{arXiv:2006.06188.} 

\bibitem{note2} Detailed discussion of the MF approximation for the 3D Bose polaron that interacts with Bose particles via the square-well potential was given in \cite{work_in_progress}.




\bibitem{Schick} M.~Schick,
\href{https://doi.org/10.1103/PhysRevA.3.1067}{ Phys.~Rev.~A {\bf 3}, 1067 (1971).}
\bibitem{Lozovik} Yu.~E.~Lozovik and V.~I.~Yudson, 
\href{https://doi.org/10.1016/0378-4371(78)90170-X}{Physica~A {\bf 93}, 493 (1978).}
\bibitem{Astrakharchik_09} G.~E.~Astrakharchik, J.~Boronat, J.~Casulleras, I.~L.~Kurbakov, and Yu.~E.~Lozovik, 
\href{https://doi.org/10.1103/PhysRevA.79.051602}{Phys.~Rev.~A {\bf 79}, 051602(R) (2009).}
\bibitem{Pastukhov_19} V. Pastukhov, 
\href{https://doi.org/10.1007/s10909-018-2082-1}{J. Low Temp. Phys. {\bf 194}, 197 (2019).}
\bibitem{Jager} J.~Jager, R.~Barnett, M.~Will, M.~Fleischhauer,
\href{https://doi.org/10.1103/PhysRevResearch.2.033142}{Phys. Rev. Research {\bf 2}, 033142 (2020).}





\end{thebibliography}
\end{document}